\documentclass[12pt]{iopart}

\usepackage{graphicx}
\usepackage{dcolumn}
\usepackage{bm}

\begin{document}


\title{Accuracy control in 
ultra-large-scale electronic structure calculation}

\author{T. Hoshi$^{1,2}$
\footnote{Present address: Department of Applied Mathematics and Physics, 
Tottori University,4-101 Koyama-Minami, Tottori,  680-8550 Japan}
}
\address{$^{1}$ Department of Applied Physics, University of Tokyo,7-3-1 Hongo, Bunkyo-ku, Tokyo 113-8656, Japan}
\address{$^{2}$ Core Research for Evolutional Science and Technology  (CREST-JST), 
Japan Science and Technology Agency, 4-1-8 Honcho, Kawaguchi-shi, 
Saitama 332-0012, Japan}
\ead{hoshi@damp.tottori-u.ac.jp}
\date{\today}

\begin{abstract}
Numerical aspects are investigated
in ultra-large-scale electronic structure calculation. 
Accuracy control methods
in process (molecular-dynamics) calculation  are focused.  
Flexible control methods are
proposed so as to control 
variational freedoms,
automatically at each time step, 
within the framework of generalized Wannier state theory.
The method is  demonstrated 
in silicon cleavage simulation with $10^2$-$10^5$ atoms.
The idea is of general importance among process calculations
and is also used in 
Krylov subspace theory,  
another large-scale-calculation theory.

\end{abstract}

\maketitle

\section{Introduction}

Nowadays one of most important scientific fields  
is process of nanostructure, 
structure in nanometer- or ten-nanometer scales, 
particularly, 
for controllability of its structure and function. 
Electronic structure calculation in these purposes
should be carried out 
with a large system ($10^3$ atoms or more) and 
a meaningful timescale.
For a decade, 
on the other hand, 
many calculation methods and related techniques 
have been proposed so as to handle such  large systems. 
See reviews 
 \cite{REVIEW-ON, REVIEW-ON2} 
and original works. 
 \cite{ MAURI, LNV1993, ORDEJON1993, 
Goedecker1994, KOHN96, HOSHI1997, 
Roche-Mayou, HOSHI2000A, Ozaki-Terakura2001, TSURUTA2001, 
SIESTA, Bowler, 
HOSHI2003A, TAKAYAMA2004A, HOSHI2005A, ONETEP, 
TAKAYAMA2006, HOSHI2005B, HOSHI2006A, OZAKI2006}
In these methodologies, 
one-body density matrix 
or Green's function is calculated, instead of one-electron eigenstates 
and  the calculation is carried out with real-space representation.
A  physical quantity $\langle X \rangle $ is given as a trace form
\begin{equation}
\langle X \rangle 
 = {\rm Tr}[ \rho X ]
  = \int\int d{\bm r}d{\bm r}^\prime 
  \rho ({\bm r},{\bm r}^\prime ) X({\bm r}^\prime, {\bm r}).
\label{TRACE-EQ}
\end{equation}
with the density matrix $\rho$. 
If the matrix $X({\bm r},{\bm r}^\prime)$ is of short range,
the off-diagonal long-range component of the density matrix 
does not contribute to the physical quantity $\langle X \rangle $,
which is crucial for practical success of large-scale calculations. 
\cite{KOHN96}

\begin{figure}[tbh]
\begin{center}
  \includegraphics[width=12cm]{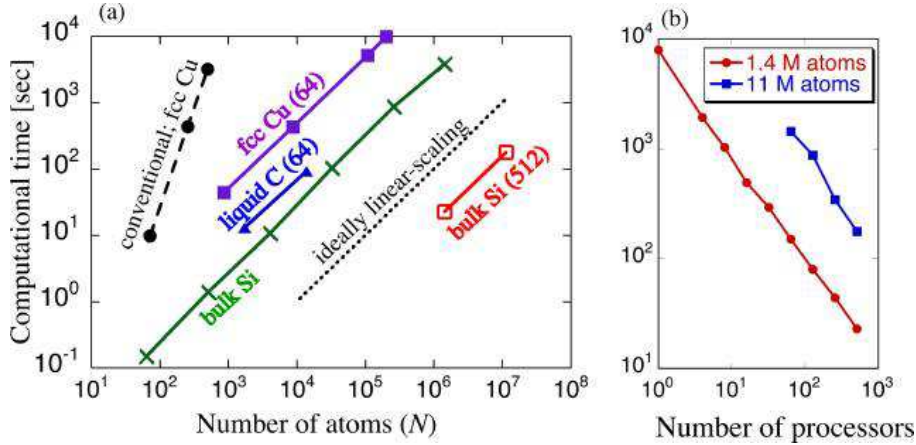}
\end{center}
\caption{
(color online) 
Computational time of 
the ultra-large-scale calculation with up to 11,315,021 atoms.
The time of our methods are plotted
for fcc Cu, liquid C and  bulk Si. 
Optimal solver method is chosen
for each system; \cite{HOSHI2006A}
Wannier state theory in perturbative procedure is chosen 
for bulk silicon and Krylov subspace theory is chosen 
for other cases. 
(a)  Comparison among materials. 
As a reference data, 
the time of the conventional eigenstate calculation 
is also plotted for fcc Cu with single CPU.
See Refs. 
\cite{HOSHI2003A, HOSHI2005A,HOSHI2006A}
for details. 
The computations were carried out using Intel or SGI CPUs.
In parallel computation, 
the number of processors (CPU cores) is indicated 
inside the parenthesis. 
(b) The time of bulk silicon 
with 1,423,909 atoms and 11,315,021 atoms,
 is measured 
as a function of processors using SGI origin 3800$^{\rm TM}$. 
\label{FIG-BENCH}
}
\end{figure}%

As ones of these works, 
we have developed a set of theories and program codes
and applied them to silicon, carbon and metal systems 
with Slater-Koster-form Hamiltonians. 
\cite{HOSHI2003A, TAKAYAMA2004A, HOSHI2005A,TAKAYAMA2006, HOSHI2005B, HOSHI2006A}
These theories are constructed from 
fundamental theory of generalized Wannier state or Krylov subspace.
An overview is given in Ref.~\cite{HOSHI2006A}. 
Figure \ref{FIG-BENCH} demonstrates our methods with or without parallel computers, 
in which the computational cost is 
\lq order-$N$' or linearly proportional to the system size ($N$), 
up to ten-million atoms 
and shows a satisfactory performance 
in parallel computation. 
We note that the electronic property,
such as density of states, is also calculated. 
~\cite{TAKAYAMA2004A, TAKAYAMA2006, HOSHI2006A}

These large-scale-calculation methods 
have controlling parameters for calculating electronic freedoms,
which gives accuracy and computational cost.
In the present paper, 
we will introduce flexible methods
of controlling electronic freedoms for optimal computational cost,
and will be demonstrated  
within the framework of generalized Wannier state theory.
The methods are crucial, particularly,  
in a dynamical process or 
a molecular dynamics (MD) calculation.
This paper is organized as follows;
An overview of theory and 
example of silicon cleavage are given in 
Sec~\ref{SEC-THEORY}. 
Then the flexible control methods 
are introduced 
and demonstrated 
in Sec.~\ref{SEC-FRAC-TECH}. 
We point out that 
similar flexible control methods 
are used in Krylov subspace theory.
Section \ref{SUMMARY} is devoted to 
concluding remarks.

\begin{figure}[hbt]
\begin{center}
  \includegraphics[width=13cm]{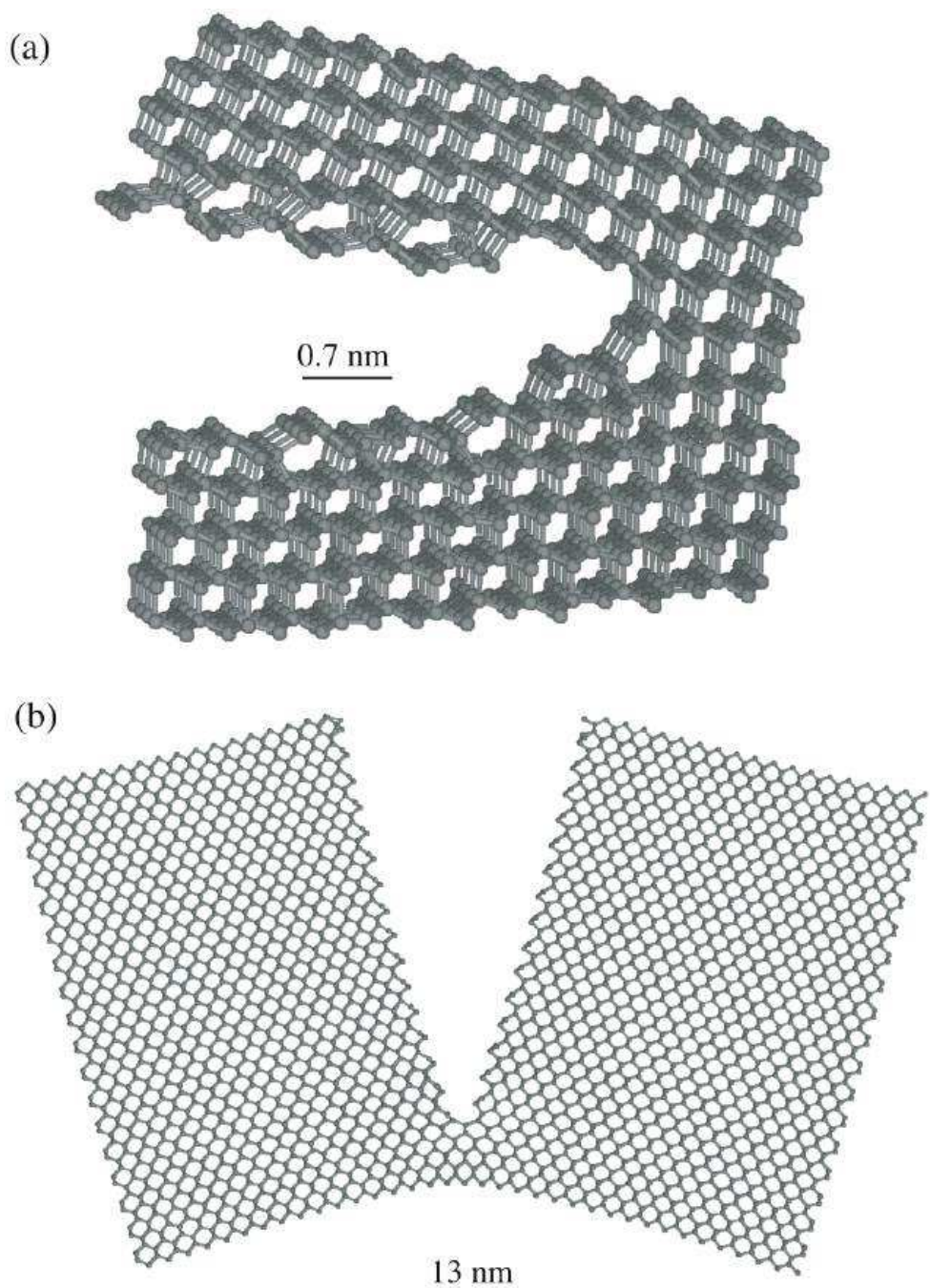}
\end{center}
\caption{
(color online) 
Simulation results of silicon cleavage. 
(a) A sample with 1,112 atoms.
The resultant cleaved surface 
shows a (111)-2$\times$1 reconstruction  and 
contains a step structure.
(b) A sample with 10,368 atoms. 
The resultant cleaved surface shows
a buckled (110) reconstruction.  
\label{fig-Si110}
}
\end{figure}

\section{Theoretical overview and examples}
\label{SEC-THEORY}

The calculation in this paper was carried out 
in the theoretical framework of generalized Wannier state.
\cite{KOHN-WANI73,KOHN-WANI93,
MAURI, ORDEJON1993, MARZARI, HOSHI2000A, HOSHI2001A, ANDERSEN2003, GESHI,THESIS}
A physical picture of the generalized Wannier states $\{  \phi_i^{\rm (WS)}  \}$
is localized chemical wave function in condensed matters,
such as a bonding orbital or a lone-pair orbital
with a slight spatial extension or \lq tail'.
The suffix $i$ of a wavefunction $\phi_i^{\rm (WS)}$ 
indicates the position of its localization center, such as bond site.
Their wavefunctions $\{  \phi_i^{\rm (WS)}  \}$
are equivalent to the unitary transformation of occupied eigen states 
and the density matrix is given as
\begin{eqnarray}
 \rho(\bm{r}, \bm{r}') = 
 \sum_{j=1}^{\rm occ.} \phi_j^{\rm (WS)}(\bm{r}),  \, 
 \phi_j^{\rm (WS)}(\bm{r}')
 \label{DM-WS}
\end{eqnarray}
where wavefunctions are described as real number.  
The Wannier state theory is suitable for large systems,
particularly,
when a dominant number of wavefunctions are well localized.
The present calculations were carried out 
by a variational procedure 
\cite{HOSHI2000A, HOSHI2006A,THESIS}

Hereafter silicon cleavage process is calculated
with a transferable Hamiltonian in the Slater-Koster form \cite{KWON}. 
Nanometer-scale or ten-nano-meter-scale samples are cleaved under external load.
Figure \ref{fig-Si110} shows examples of the resultant cleaved samples 
that contain experimentally observed cleavage planes,
(111) and (110) planes; 
In Fig, \ref{fig-Si110}(a), the resultant sample contains 
a cleaved  Si(111)-2$\times$1 
surface. \cite{HOSHI2005A} 
A pair of five-and seven-membered rings 
appears in the cleavage propagation direction, 
[$2\bar{1}\bar{1}$] direction, 
which forms the unit cell of 
the 2$\times$1 structure called Pandey structure. 
\cite{PANDEY,SI111-21-PARRINELLO-MD,SI111-HYB}
As an interesting feature of the present result,
the  cleaved surface contains 
a step structure with a six-membered ring at the step edge,
which is compared to experiments. \cite{HOSHI2005A}
As details, 
the sample consists of 1,112 atoms and 
the periodic boundary condition is imposed,
by eight atomic layers,  
in the direction  perpendicular to
the cleavage propagation direction.
In the present case,
an additional periodicity, by two atomic layers,
is imposed as a constraint on the atomic structure.
We note that the 2x1 structure appears even without 
the additional periodicity. 
See papers \cite{HOSHI2005A} for more details and 
results of larger samples with $10^5$ atoms. 
In Fig, \ref{fig-Si110}(b), the resultant cleaved 
surface is a buckled (110) surface that
appears in  textbooks in surface physics or papers 
such as Refs. \cite{PEREZ, STEKOL}. 
As details, 
the sample consists of  10,368 atoms and 
the periodic boundary condition is imposed,
by eight atomic layers,  
in the direction  perpendicular to
the cleavage propagation direction.
The physical discussions in cleavage dynamics
are found in Refs. \cite{HOSHI2003A,HOSHI2005A, HOSHI2006A} 
and reference therein.


\begin{figure}[hbt]
\begin{center}
  \includegraphics[width=13cm]{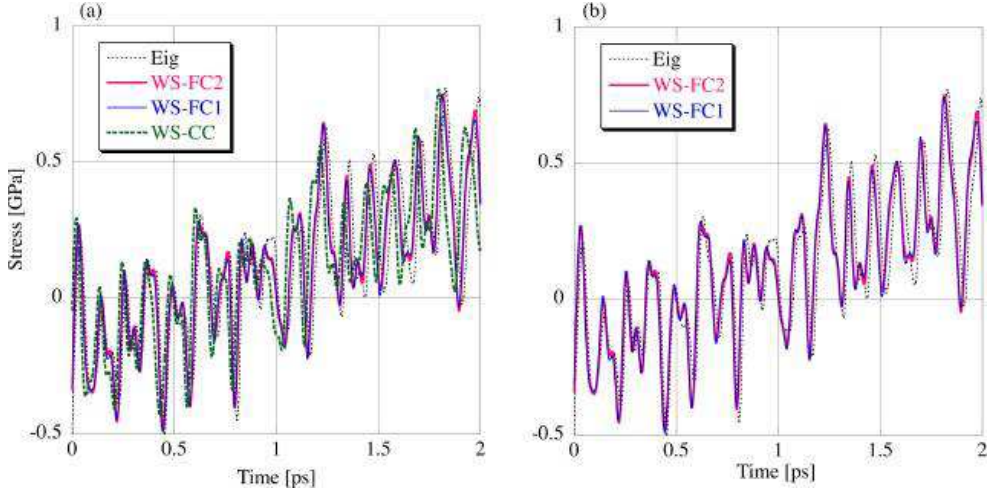}
\end{center}
\caption{
(color online) 
(a) The stress value of nanocrystalline silicon of 91 atoms
with thermal vibration, 
in which the sample is stretched 
by a [001] uniaxial load.
The calculations were carried out by 
the Wannier state method with
(i) \lq constant cutoff' (WS-CC),
(ii) \lq flexible control at the first level' (WS-FC1), and
(iii) \lq flexible control at the second level' (WS-FC2). 
See text for explanation. 
The conventional eigen state method (Eig) 
was also carried out as a reference data.
(b) The same data set as in (a) are plotted but 
the data of the CC method is ignored,
so as to clarify
a significantly better agreement among the other methods. 
\label{fig-on-dg-91atom-all}
}
\end{figure}

\section{Flexible methods for accuracy control \label{SEC-FRAC-TECH}}

\subsection{Three methods in Wannier state theory}

Here we describe the accuracy control methods 
\cite{HOSHI2003A,HOSHI2005A,THESIS}
used in the above MD calculation. 
In generalized Wannier state theory, 
the region for localization constraint
for each Wannier state
is variational freedoms
that governs accuracy and  computational cost.
Therefore we will concentrate the methods of  
setting the localization region for each wavefunction
at each time step. 

Here the three methods of accuracy control 
in the Wannier state calculation are proposed. 
Among all the methods, 
the localization constraint 
on each wavefunction $\phi_i^{\rm (WS)}$ 
is imposed as a spherical region whose
center is the weighted center of the wavefunction
$\bm{r}_i^{(\rm WS)} \equiv 
\langle \phi_i^{\rm (WS)} | \hat{\bm{r}} | \phi_i^{\rm (WS)} \rangle$.
Therefore, the cutoff radius of the spherical region,
denoted $R_i^{\rm (WS)}$,
 mainly contributes to accuracy.
We also 
denote $N_i^{\rm (WS)}$ 
as the number of atoms inside the localization region of 
the $i$-th Wannier state. 
Three methods for determination of 
the radius are used;
(i) \lq constant cutoff' method (WS-CC method)
(ii) \lq flexible control method at the first level' (WS-FC1 method), and  
(iii) \lq flexible control method at the second level' (WS-FC2 method). 
See below for explanation of these methods. 

The method is demonstrated 
in nanocrystalline silicon of an isolated cubic sample with 91 atoms. 
\cite{THESIS}
The sample is thermally vibrated with 300 K and 
an additional slow constant-velocity motion is introduced 
for the atoms on the sample surface. 
As a result, 
the sample is stretched in the [001] direction with thermal vibration. 
Figure \ref{fig-on-dg-91atom-all} 
shows the trajectory of calculated stress $\sigma$. 
In Fig.\ref{fig-on-dg-91atom-all}(a), 
the results of 
three controlling methods for Wannier states 
are compared. 
Figure \ref{fig-on-dg-91atom-all}(a) also contains
a result of conventional eigen-state calculation 
as a reference data, 
in which the temperature (level-broadening) parameter 
of $\tau=0.1$ eV is used for electronic system. 

In the WS-CC method of Fig.~\ref{fig-on-dg-91atom-all}, 
the radius is chosen to be a constant value 
of $R_i^{\rm (WS)}= 2.5 d_0$, 
where $d_0 (=2.35 {\rm \AA})$ is the equilibrium bond length. 
This value is chosen 
for all Wannier states through the simulation. 
Without an external load, 
this radius sets the localization region of the Wannier states
to about $N_i^{\rm WS}=40$ atoms.
We should say  that a results with the CC method 
is expected to be rather poor, 
because  
the sample in the present MD simulation will be 
stretched by the external load and
the number of atoms within the localization region
tends to decrease during the MD simulation.
This point will be confirmed numerically in the last paragraph of the present subsection.

A better way for accuracy control is 
to give the number of atoms in the localization region, 
$N_i^{\rm WS}$, instead of a given radius $R_i^{\rm (WS)}$,
which realized a flexible control for the localization radius.
In this method,
the radius $R_i^{\rm (WS)}$ is chosen so that 
the localization region contains
a given number, $N_i^{\rm (WS, min)}$, of atoms or more.
This method is called flexible control method at the first level 
(WS-FC1 method). 
In Fig.~\ref{fig-on-dg-91atom-all}(a) 
we choose the value of $N_i^{\rm (WS, min)}=40$ .
In results, the localization radius $R_i^{\rm (WS)}$ 
may be different among the Wannier states and
the number of atoms within the localization region ($N_i^{\rm WS}$)
always satisfies $N_i^{\rm (WS)} \ge N_i^{\rm (WS,min)} = 40$.

Now we explain the third method for setting the localization region, 
called flexible control method at the second level (WS-FC2 method).
In the program code, an iterative solution procedure is carried out
for an equation of generalized Wannier states.
See Refs.~\cite{HOSHI2000A, HOSHI2006A, THESIS} 
for the explicit expression of the equation. 
Since the residual of the equation ($\delta \phi_i$) 
is well defined for each wavefunction $\phi_i^{\rm (WS)}$, 
the accuracy of a calculated wavefuncion can be rigorously monitored by 
the residual norm $|\delta \phi_i|$. 
The residual norm vanishes, when the calculated wavefuncion will be exact
($|\delta \phi_i| \rightarrow 0$). 
When the wavefunction $\phi_i$ has 
a large residual norm $|\delta \phi_i|$,
a larger number of atoms ($N_i^{\rm (WS)}$) 
should be assigned  inside the localization region
so as to reduce the residual norm $|\delta \phi_i|$.
In the present code, 
the assignment is carried out automalically for each wavefunction
at each time step. 
In the calculation with Fig.~\ref{fig-on-dg-91atom-all}(a), 
we classify all wavefunction into three classes 
with different numbers $N_i^{\rm (WS, min)}$; 
$N_i^{\rm (WS, min)} = 40, 60$ or  80. 
The classification procedure is carried out  
with the averaged value $\delta \phi_{\rm av}$  of the residual norm 
among all wavefunctions 
$\{  | \delta \phi_i | \}$ .
If the residual norm of a wavefunction ($| \delta \phi_i |$)
is almost the same as its averaged value
($| \delta \phi_i | \le 1.2 \delta \phi_{\rm av}$), 
the number of  $N_i^{\rm (WS, min)}$ is set to be
the small one ($N_i^{\rm (WS, min)} = 40$).
If the residual norm of a wavefunction ($| \delta \phi_i |$)
is slightly larger than its averaged value
($ 1.2 \delta \phi_{\rm av} \le | \delta \phi_i | \le 1.5 \delta \phi_{\rm av}$), 
the number of  $N_i^{\rm (WS, min)}$ is set to be
the middle one ($N_i^{\rm (WS, min)} = 60$).
If the residual norm of a wavefunction ($| \delta \phi_i |$)
is larger than 150 \% of its averaged value
($ 1.5 \delta \phi_{\rm av} \le | \delta \phi_i | $), 
the number of  $N_i^{\rm (WS, min)}$ is set to be
the large one ($N_i^{\rm (WS, min)} = 80$).

When 
the three control methods with Wannier states,
WS-CC, WS-FC1 and WS-FC2 methods, 
are compared, in  
Fig.~\ref{fig-on-dg-91atom-all}(a),
with a reference data by
the conventional eigen state method,
one finds that
the flexible control methods, 
FC1 and  FC2 methods, 
are siginificantly better in accuracy than the CC method. 
This statement is clarified in Fig.~\ref{fig-on-dg-91atom-all}(b), 
when the trajectories without that of the CC method
show a better agreement.

\subsection{Discussions}

Although the flexible control methods give, in general,  
a better accuracy during the MD simulation 
than the CC method, 
any of the three methods, WS-CC, WS-FC1 and WS-FC2 methods,  
is sufficient for discussing physical quantities in the present case;
For example, 
the averaged gradient of Fig.~\ref{fig-on-dg-91atom-all} 
is proportional to the Young modulus 
in the [001] direction $(E_{100})$,
because the stretching motion is realized within a constant velocity.
The Young modulus is estimated, 
commonly among four calculation methods, 
to be $E_{100} \approx 100$GPa, 
where the estimated value may include
an error on the order of 10 \%. 
The estimated value 
is comparable with the experimental bulk value
$E_{100} \approx 130$ GPa but is deviated, 
owing to the small system size. 
Satisfactory results are given also 
for critical stress for cleavage; 
$\sigma_{\rm c} =2.5 - 3.0$ GPa.
Moreover the cleavage propagation velocity (not shown)
agrees well among the three Wannier state calculations
and  the eigen-state calculation.
Note that the discussion of these quantities 
in nanocrystalline silicon 
is given in Ref.~\cite{HOSHI2003A}. 
 
The WS-FC2 method is required in several simulations and
one example is the case of 
Fig.~\ref{fig-Si110}(a) or 
silicon cleavage with Si(111)-2$\times$1 cleaved surface;
The elementary reconstruction process 
occurs among several bond sites 
including surface and subsurface layers, 
\cite{HOSHI2005A}
and a larger region is required for describing wavefunctions 
near the cleaved surface. 
Since the number of wavefunctions near the cleaved surface 
accounts for only a small fraction, typically 10 \%, 
of the total number of wavefunctions,
the total computational cost of the FC2 method is moderate,
when compared with the CC method.  

Finally we note that a similar flexible control is also used in 
Krylov-subspace theory,
another theory for large-scale-calculation theory.
See Appendix of Ref. \cite{HOSHI2006A}.

\section{Concluding remarks \label{SUMMARY}}

In this paper, 
we focus the way of accuracy control 
in dynamical process or MD simulation.
Flexible control methods are proposed so as to realize
large-scale process (MD) calculation,
in which the electronic freedoms 
are determined optimally at each time step. 

Since nature (or electronic structure) of physical system 
can change during a dynamical process, 
flexible control methods proposed here 
are crucial, generally, among large-scale calculations,
when one would like to achieve 
a proper balance between accuracy and computational cost.

\ack

This work is supported by a Grant-in-Aid from
the Ministry of Education, Science, Sports and Culture of Japan.
Numerical calculation was partly carried out 
in the Japan Atomic Energy Research Institute, 
the Institute for Solid State Physics, University of Tokyo,
and the Research Center for Computational Science, Okazaki.



\newcommand{\noop}[1]{}

\end{document}